\documentclass[10pt,a4paper]{article}

\usepackage[english]{babel}
\usepackage{authblk}
\usepackage{graphicx}
\usepackage{hyperref} 
\usepackage[left=1in,right=1in,top=1in, bottom=1in]{geometry}
\usepackage{amsmath}
\usepackage{amssymb}
\usepackage{amsfonts}
\usepackage{multicol}
\usepackage{slashed}
\usepackage{float}
\usepackage{caption}
\usepackage{subcaption}
\usepackage{cite}
\usepackage{amsthm}
\usepackage{xcolor}
\usepackage{musicography}
\usepackage{bbm}
\usepackage{enumitem} 
\usepackage{csquotes} 

\newcommand{\ri}{{\rm i}}
\newcommand{\M}{\mathcal{M}}
\newcommand{\R}{\mathbb{R}}
\newcommand{\E}{\mathbb{E}}
\newcommand{\p}{\partial}

\def\XXint#1#2#3{{\setbox0=\hbox{$#1{#2#3}{\int}$ }
		\vcenter{\hbox{$#2#3$ }}\kern-.6\wd0}}

\title{\bf Quantum Mechanics from Stochastic Processes}
\author[1]{Folkert~Kuipers\thanks{E-mail: Kuipers@na.infn.it}}
\affil[1]{\em INFN, Sezione di Napoli\authorcr \em Complesso Universitario di Monte S. Angelo\authorcr \em Via Cintia Edificio 6, 80126 Napoli, Italy}

\begin{document}

\maketitle

\begin{abstract}
We construct an explicit one-to-one correspondence between non-relativistic stochastic processes and solutions of the Schr\"odinger equation and between relativistic stochastic processes and solutions of the Klein-Gordon equation. The existence of this equivalence suggests that the Lorentzian path integral can be defined as an It\^o integral, similar to the definition of the Euclidean path integral in terms of the Wiener integral. Moreover, the result implies a stochastic interpretation of quantum theories.
\end{abstract}

\begin{multicols}{2}

\section{Introduction}

Ever since the introduction of quantum mechanics, its foundations have puzzled the minds of many physicists, as this theory introduces various notions that were not encountered previously in classical physics. However, over the course of the last century, it has become clear that many features that were initially thought to be puzzling and unique to quantum theories have analogues in classical stochastic theories.
\par 

Indeed, it is now well established that the concept of phase-space non-commutativity, encountered in quantum mechanics, is a generic feature of stochastic theories. In particular, in the theory of Brownian motion, it can be regarded as an immediate consequence of the non-differentiability of the paths of a Brownian motion \cite{Feynman:1948ur}, and can be derived from the definition of the It\^o integral along a Wiener process \cite{Ito}.
\par 

Directly related to this is the presence of uncertainty principles in both quantum mechanics and statistical physics. An uncertainty relation between energy and temperature in statistical physics was for example obtained in Ref.~\cite{Mandelbrot}. Moreover, position-momentum uncertainty relations have been shown to be a generic feature of stochastic theories, cf. e.g. Ref.~\cite{Koide:2012ya}. 
\par 

Furthermore, the Fock space representation that plays a prominent role in the study of quantum theories, has also been constructed for the Wiener process describing Brownian motion, cf. e.g. Ref.~\cite{Biane:2010}. In this context, the existence of the Fock space is a consequence of the predictable representation property of the Wiener process, which allows to calculate the expectation of any observable of the Brownian theory using a Wiener-It\^o expansion (a.k.a. Wiener chaos \cite{Wiener:1923}).
\par 

Perhaps, the most striking similarity between quantum mechanics and Brownian motion is the fact that the time evolution of the probability density in both theories is governed by a diffusion equation, i.e. an equation of the form
\begin{equation}\label{eq:DiffEq}
	\alpha \, \frac{\p}{\p t} \, \Psi(x,t) = H(x,\hat{p},t) \, \Psi(x,t) \, ,
\end{equation}
where $\alpha=|\alpha|\, e^{\ri \, \phi}\in\mathbb{C}$, $t\in\mathcal{T}=[0,T]\subset\R$ and $x\in\M$. Furthermore, $H$ is a Hamiltonian or diffusion operator that is second order in the operator $\hat{p}$, acting on complex valued funtions $\Psi:\M\times\mathcal{T}\rightarrow \mathbb{C}$. When $\alpha\in\R$, the diffusion equation is called the heat equation\footnote{For $\alpha\in(0,\infty)$ one obtains the heat equation and for $\alpha\in(-\infty,0)$ the time-reversed heat equation.} and describes the dynamics of the probability density for a Brownian motion, while for $\alpha\in\ri\times\R$, the diffusion equation is called the Schr\"odinger equation\footnote{For $\alpha\in\ri\times(0,\infty)$ one obtains the Schr\"odinger equation and for $\alpha\in\ri \times (-\infty,0)$ the time-reversed Schr\"odinger equation.} and describes the dynamics of the probability density in quantum mechanics.
\par 

Let us now focus on the simple case, where the configuration space is the Euclidean space\footnote{Starting in section \ref{eq:RLT}, we will generalize this to pseudo-Riemannian manifolds $(\M,g)$.} $\R^n$ and $\hat{p}$ is a differential operator given by
\begin{equation}\label{eq:CanMomentum}
	\hat{p} = - \alpha \, \frac{\p}{\p x} \, .
\end{equation}
In addition, we consider the Hamiltonian
\begin{equation}
	H = \frac{\delta^{ij}}{2\,m}  \left(\alpha \, \frac{\p}{\p x^i} + q A_i \right) \left( \alpha \, \frac{\p}{\p x^i} + q A_j \right) + \mathfrak{U}
\end{equation}
with $m$ the mass of the particle, $q$ its charge under the covector potential $A_i(x,t)$ and $\mathfrak{U}(x,t)$ a scalar potential.
\par 

Using the diffusion equation and this definition of momentum, one can derive various features that are shared by Brownian motion and quantum mechanics. The first is that the evolution equation \eqref{eq:DiffEq} implies the presence of a superposition principle, such that for time-independent potentials bound states can be decomposed as
\begin{equation}
	\Psi(x,t) = \sum_k c_k \, \psi_k(x) \, e^{\frac{E_k}{\alpha} t} \, .
\end{equation}
The second is that the momentum operator \eqref{eq:CanMomentum} implies the canonical commutation relation
\begin{equation}
	[\hat{x}^i,\hat{p}_j] = \alpha \, \delta^i_j \, .
\end{equation}
The third is the presence of an uncertainty principle between position and momentum of the form
\begin{equation}
	\sqrt{{\rm Var}(X^i) \, {\rm Var}(P_j)} \geq \frac{|\alpha|}{2} \left( 1 + \cos \phi \right) \delta^i_j \, ,
\end{equation}
where ${\rm Var}(X)$ denotes the variance of $X$.
\par 

The similarities between the heat equation and the Schr\"odinger equation have led many scientists to speculate that quantum mechanics can be understood as a Brownian diffusion process with a complex diffusion constant, cf. e.g. \cite{Gelfand}. However, the realization of this suggestion suffers from serious difficulties: for real diffusion constants, the Feynman-Kac theorem \cite{FKac} establishes an equivalence between the heat equation and the theory of Brownian motion, whereas such an equivalence cannot be established straightforwardly for $\alpha\notin \R$ \cite{Cameron,Daletskii}. 
\par 

Despite these difficulties, the similarity between the heat equation and the Schr\"odinger equation has been put to good use in the Euclidean approach, cf. e.g. Ref.~\cite{Glimm:1987ng}, where a quantum theory is mapped onto a statistical theory using a Wick rotation \cite{Wick:1954eu}. This mapping allows to derive features of the quantum theory by studying the corresponding statistical theory.\footnote{The statistical theory is often referred to as the Euclidean theory as the Wick rotation does not only change the value of the diffusion constant, but also the signature of the spacetime.}
\par 

In addition, there have been various attempts to generalize the Wiener process, such that an equivalence with the Schr\"odinger equation can be obtained. The most advanced of these attempts is the theory of stochastic mechanics \cite{Fenyes,Nelson,Nelson:1967,Nelson:1966sp,Guerra:1981ie,Pavon:1995April,Pavon:2000}.
\par 

In this letter, we build on this theory, and show, by a complexification of the stochastic noise, that for any $\alpha\in\mathbb{C}$, there exists a stochastic process whose equations of motion are equivalent to the diffusion equation \eqref{eq:DiffEq}. Moreover, we provide an explicit extension to relativistic theories and theories on curved space(time). 
For a detailed account of the various results, we refer to Ref.~\cite{Kuipers:2023pzm}.
\par

Throughout the letter, we use natural units such that $\hbar=c=G=k_B=1$ and a $(-+++)$ metric signature.

\section{The Wiener process}
Let us start by considering a macroscopic particle on the Euclidean space $\M=\R^n$ with mass $m$ and energy $E_0$. We assume that this particle is submerged in a fluid of a large number $N$ of microscopic particles, with energies $\{E_i:i\in\{1,...,N\}\}$, that are continuously moving around and interacting with each other. Due to this thermal motion, some of the microscopic particles will hit the macroscopic particle and exchange momentum with this particle. Consequently, the macroscopic particle is subjected to a large number of small kicks at discrete times $\{t_1, t_2 , ...\}\subset\mathcal{T}$, and will display the behavior of a Brownian motion.
\par 

Let us now take the continuum limit of this Brownian motion. More precisely, we take the limit of the number of microscopic particles $N\rightarrow\infty$ and their respective energy $E_i\rightarrow0$ (for $i\geq1$) such that the total energy $\sum_{i=0}^N E_i$ of the system (macroscopic particle) and background (microscopic particles) is conserved. In this continuum limit, the macroscopic particle is kicked at every time $t$ in the continuous time interval $\mathcal{T}$ and an infinitesimal amount of energy is exchanged at every kick. 
\par

This continuum limit of Brownian motion is described by the Wiener process, which is defined as a stochastic process $M:\mathcal{T} \times \Omega \rightarrow \R^n$ satisfying the following properties:
\begin{itemize}
	\item almost sure continuity: $\forall \, t\in\mathcal{T}$, \\ $\mathbb{P}\big(\lim_{s\rightarrow t} ||M_s - M_t|| = 0\big)=1\,$; 
	\item independent increments: $\forall \, t_1<...<t_4 \in \mathcal{T}$, $\mathbb{P}\big(M_{t_4} - M_{t_3}\, \big|\, M_{t_2} - M_{t_1} \big) = \mathbb{P}\big(M_{t_4} - M_{t_3} \big)\,$;
	\item Gaussian increments: $\forall \, t_1<t_2\in\mathcal{T},$\\
	$(M_{t_2}-M_{t_1})\sim \mathcal{N}(0,\frac{\alpha}{m} \, \delta^{ij} \, (t_2-t_1))\,$,\\ where ${\alpha\geq0}$ is the diffusion constant.
\end{itemize}
For every $\omega\in\Omega$, $M(\omega):\mathcal{T}\rightarrow\R^n$ defines a sample path of the process, and $\mathbb{P}:\Sigma(\Omega)\rightarrow[0,1]$, with $\Sigma(\Omega)$ a sigma algebra over the sample space $\Omega$, is a probability measure that assigns a probability to every sample path.
\par 

Although this standard definition of the Wiener process is very intuitive and practical in numerical studies, there exists another definition of the Wiener process that is more useful for analytical purposes. In this equivalent definition, known as the L\'evy characterization \cite{Levy}, the Wiener process is defined by the following properties:
\begin{itemize}
	\item martingale property: $\forall \, t_1<t_2\in\mathcal{T},$\\
	$\E\big[M_{t_2} \, \big| \, \{M_s:s\in[0,t_1]\}\big] = M_{t_1}\,$;
	\item structure relation:
	$\forall \ t\in\mathcal{T},$\\ $d[M^i,M^j]_t = \frac{\alpha}{m} \, \delta^{ij} \, dt\,$,
\end{itemize}
where the expectation value $\E$ is defined by the Lebesgue integral
\begin{equation}
	\E[M] = \int_\Omega M(\omega) \, d\mathbb{P}(\omega) \, ,
\end{equation}
and the bracket $d[.,.]$ is called the quadratic variation. In differential notation, it is given by
\begin{equation}\label{eq:DiffObject2}
	d[M,M]_t = (M_{t+dt} - M_t) \otimes (M_{t+dt} - M_t)\, .
\end{equation}

\section{Processes with drift}
The Wiener process is a martingale, i.e. a driftless stochastic process. We will now add drift to this process by imposing that the trajectory of the macroscopic particle can be decomposed as
\begin{equation}\label{eq:DoobMeyer}
	X^i_t = C^i_t + e^i_a(X_t) \, M^a_t \, ,
\end{equation}
where $C^i_t$ is a continuous deterministic trajectory and $M^a_t$ a Wiener process. Such a process is called a semi-martingale. Here, we will assume that $e^i_a(X_t)$ is a polyad (vielbein) that defines a frame at every point $x\in\M$. Thus, the noise $M^a_t$ is defined on the reference frame $F=\R^n$ that is attached to every point $x\in\M$, whereas $X$ and $C$ are defined on the configuration space $\M$. On the flat space $\M=\R^n$, the polyad is simply given by $e^i_a(X_t)=\delta^i_a$.
\par 

The dynamics of this process is governed by a stochastic differential equation of the form
\begin{equation}\label{eq:Ito}
	\begin{cases}
		dX^i_t &= v^i(X_t,t) \, dt + \delta^i_a \, dM^a_t \, , \\
		d[M^a,M^b]_t &= \frac{\alpha}{m} \, \delta^{ab} \, dt \, ,
	\end{cases}
\end{equation}
where the velocity field is defined by the conditional expectation
\begin{equation}
	v(X_t,t) = \lim_{\epsilon\rightarrow 0^+} \E\left[ \frac{ X_{t+\epsilon} -X_t}{\epsilon} \, \Big| \, X_t\right]\, .
\end{equation}
We note that the first line of eq.~\eqref{eq:Ito} is equivalent to the decomposition \eqref{eq:DoobMeyer}, but reformulated as an It\^o equation. Moreover, the second line is the L\'evy characterization and ensures that $M$ is a Wiener process, such that $X$ has Gaussian noise.

\section{Two-sided processes}
At this point, the dynamics of the process $X$ is only defined with respect to the future directed time evolution. We are also interested in its past directed evolution, and impose that
\begin{enumerate}
	\item the time evolution of the process $X$ is well-defined towards both the future and past;
	\item the physical laws that govern the dynamics are invariant under time-reversal.
\end{enumerate}
These two conditions can easily be implemented, and stochastic processes that satisfy them are called two-sided processes \cite{Nelson:1967}. The two-sided Wiener process is a process that is almost surely continuous, has independent increments and has
\begin{itemize}
	\item two-sided Gaussian increments: $\forall \ t_1,t_2\in\mathcal{T},$\\
	$(M_{t_2}-M_{t_1})\sim \mathcal{N}(0,\alpha \, \delta^{ij} \, |t_2-t_1|)\,$ with $\alpha\geq0\,$.
\end{itemize}
In stochastic theories, time-reversal invariance of the laws of motion does not imply time-reversal invariance of the stochastic processes. Therefore, the decomposition \eqref{eq:DoobMeyer} changes to 
\begin{equation}\label{DoobMeyer2}
	X^i_t = C^i_{\pm,t} + e^i_a(X_t) \, M^a_t \, ,
\end{equation}
where $C_+$ defines the future directed evolution and $C_-$ the past directed evolution of $X$. Moreover, the It\^o equation \eqref{eq:Ito} is now given by
\begin{equation}\label{eq:Ito2}
	\begin{cases}
		d_\pm X^i_t &= v_\pm^i(X_t,t) \, dt + \delta^i_a \, d_\pm M^a_t \\
		d[M^a,M^b]_t &= \frac{\alpha}{m} \, \delta^{ab} \, dt \, 
	\end{cases}
\end{equation}
with It\^o velocities 
\begin{equation}\label{eq:ItoVelocity}
	v_\pm(X_t,t) = \lim_{dt\rightarrow 0} \E\left[ \frac{ d_\pm X_t}{dt} \, \Big| \, X_t\right],
\end{equation}
where
\begin{align}\label{eq:DiffObject1}
	d_+ X_t &= X_{t+dt} - X_t \nonumber\\
	d_- X_t &= X_t - X_{t-dt}
\end{align}
define the forward and backward It\^o differential. Similarly, in this differential notation, the quadratic variation is given by
\begin{align}
	d_+ [X,X]_t &= (X_{t+dt} - X_t) \otimes (X_{t+dt} - X_t) \, ,\nonumber\\
	d_- [X,X]_t &= (X_t - X_{t-dt}) \otimes (X_t - X_{t-dt}) \, ,
\end{align}
but time-reversibility of the stochastic law of motion implies
\begin{equation}
	d[X,X]_t := d_+[X,X]_t = - d_-[X,X]_t \, .
\end{equation}

\section{Stochastic calculus}
Our aim is to subject the process $X$ to various potentials and to derive equations of motion for the process. Before doing so, we will discuss some elementary notions from stochastic calculus.
\par 

Let us first recall that the differential objects $d_\pm X$ and $d[X,X]$ given in eqs.~\eqref{eq:DiffObject1} and  \eqref{eq:DiffObject2} are heuristic, but can be rigorously defined by their corresponding integral expression: for any function $f$, $f(X_t,t) \, d_\pm X$ is defined by the It\^o integrals
\begin{align}
	\int_\mathcal{T} f(X_t,t) d_+ X_t
	&=
	\lim_{N\rightarrow \infty} \sum_{k=0}^{N-1}
	f(X_{t_k},t_k) \big[X_{t_{k+1}} - X_{t_k} \big] \nonumber\\
	\int_\mathcal{T} f(X_t,t) d_- X_t
	&=
	\lim_{N\rightarrow \infty} \sum_{k=1}^{N}
	f(X_{t_k},t_k) \big[X_{t_k} - X_{t_{k-1}} \big] 
\end{align}
where $\{0=t_0<t_1<...<t_N=T\}$ is an arbitrary partition of the interval $\mathcal{T}$. Similarly, $f(X_t,t) \, d[X,X]_t$ is defined by the integral
\begin{align}
	&\int_0^T f(X_t,t) \, d[X,X]_t  = \\
	&
	\lim_{N\rightarrow \infty} \sum_{k=0}^{N-1}
	f(X_{t_k},t_k) 
	\big[X_{t_{k+1}} - X_{t_k} \big] \otimes \big[X_{t_{k+1}} - X_{t_k} \big] \nonumber
\end{align}
\par 

Now that we have provided proper definitions for the differentials $d_\pm X$ and $d[X,X]$, we can discuss the action of a differential operator $d$ on scalar functions $f(X_t,t)$, assuming that $X$ is a continuous process. In a deterministic theory, this action is given by
\begin{equation}
	df = \p_t f \, dt + \p_i f dX^i_t + O(dt^2)\,.
\end{equation}
In a stochastic theory, this differential operator $d$ is replaced by second order operators $d_\pm$ that satisfy It\^o's lemma, such that
\begin{align}
	d_\pm f &= \p_t f \, dt + \p_i f d_\pm X^i_t \nonumber\\
	&\quad \pm \frac{1}{2} \, \p_j \p_i f \, d[X^i,X^j] + o(dt)\,,\\
	d[f,g] &= \p_i f \, \p_j g \, d[X^i,X^j] + o(dt) \, .
\end{align}
An immediate consequence of this fact is that the classical Leibniz rule and chain rule, given by
\begin{align}
	d(f\, g) &= f\, dg + g \, df  \\
	d(h\circ f) &= (h'\circ f) \, df \, ,
\end{align}
are replaced by
\begin{align}
	d_\pm(f\, g) &= f\, d_\pm g + g \, d_\pm f \pm d[f,g] \\
	d_\pm(h\circ f) &= (h'\circ f) \, d_\pm f \pm \frac{1}{2} (h'' \circ f) \, d[f,f] \, .
\end{align}
\par

Finally, we note that for a process $X$ that can be decomposed into a deterministic drift and a noise, as in eq.~\eqref{eq:DoobMeyer}, the quadratic variation is given by
\begin{equation}
	d[X^i, X^j]_t = e^i_a e^j_b \, d[M^a,M^b] + o(dt) \, ,
\end{equation}
which follows from the fact that $dC=O(dt)$, whereas $dM=o(1)$.

\section{Stochastic Lagrangian}
We will subject the process $X$ to a classical Lagrangian of the form
\begin{equation}\label{eq:ClassLag}
	L(x,v,t) = \frac{m}{2} \, \delta_{ij} v^i v^j + q \, A_i(x,t)\, v^i - \mathfrak{U}(x,t) \, .
\end{equation}
This classical Lagrangian is gauge invariant for deterministic trajectories. Indeed, if we add a total derivative $\int dF(x,t)$ to the action $S=\int Ldt$, the new action $\tilde{S}= S + \int dF$ can be written as $\tilde{S}=\int\tilde{L}dt$, where the new Lagrangian $\tilde{L}$ depends on the potentials $\tilde{A}_i=A_i + q^{-1} \p_i F$ and $\tilde{\mathfrak{U}}=\mathfrak{U} - \p_t F$. 
\par 

For stochastic trajectories a total derivative is given by $\int d_\pm F$, which generates second order derivatives $\p_j\p_iF$. Therefore, the Lagrangian \eqref{eq:ClassLag} does not respect gauge invariance, if it describes a stochastic theory. This issue can be resolved by constructing a gauge invariant stochastic Lagrangian that is given by \cite{Kuipers:2023pzm}
\begin{equation}\label{eq:ItoLag}
	L^\pm(x,v_\pm,v_2,t) = L_0^\pm(x,v_\pm,v_2,t) \pm L_\infty(x,v_\circ)
\end{equation}
with finite Lagrangian
\begin{equation}\label{eq:FinLag}
	L_0^\pm = \frac{m}{2} \, \delta_{ij} v_\pm^i v_\pm^j + q \, A_i v_\pm^i \pm \frac{q}{2}  v_2^{ij} \p_j A_i - \mathfrak{U}
\end{equation}
and a divergent part that is defined by the integral condition\footnote{The presence of this divergent term resolves Wallstrom's criticism of stochastic mechanics \cite{Wallstrom:1988zf,WallstromII}.}
\begin{equation}\label{eq:divLag}
	\E\left[ \int L_{\infty} dt \right] = \E\left[\int \frac{m}{2} \delta_{ij} \, d[x^i,v_\circ^j] \right].
\end{equation}
In this expression, the second order velocity is associated to the quadratic variation, such that
\begin{align}
	v_2^{ij}(X_t,t) 
	&= \lim_{dt\rightarrow 0}\E\left[\frac{d[X^i,X^j]}{dt}\, \Big|\,X_t\right] \nonumber\\
	&=  \frac{\alpha}{m} \, \delta^{ij} 
\end{align}
and the Stratonovich velocity is given by
\begin{equation}
	v_\circ = \frac{1}{2} \, (v_+ + v_-)\, .
\end{equation}

\section{Equations of Motion}\label{sec:EQM}
Starting from the stochastic Lagrangian, one can use stochastic variational calculus to derive the stochastic Euler-Lagrange equations, the stochastic Hamilton equations and the stochastic Hamilton-Jacobi equation, cf. e.g. Refs.~\cite{Zambrini,Huang:2022,Kuipers:2023pzm}. 
\par

Here, we focus on the Hamilton-Jacobi equations. These are given by \cite{Kuipers:2023pzm}
\begin{equation}\label{eq:HamJac}
	\begin{cases}
		\p_i S^\pm(x,t) &= p^\pm_i \, , \\
		\p_t S^\pm(x,t) &= - H^\pm_0(x,p^\pm,\p p^\pm,t)\, ,
	\end{cases}
\end{equation}
where $S^\pm$ is Hamilton's principal function associated to $L^\pm$, the momentum is defined by
\begin{equation}\label{eq:Momentum}
	p^\pm(x,t) = \frac{\p L_0^\pm}{\p v_\pm} 
	= m \, \delta_{ij}\,v_\pm^j + q \, A_i(x,t)
\end{equation}
and the Hamiltonian by the second order Legendre transform \cite{Huang:2022}
\begin{align}
	H_0^\pm &= p^\pm_i v_\pm^i \pm \frac{1}{2} \, \p_j p_i^\pm v_2^{ij} - L_0^\pm \nonumber\\
	&= \frac{m}{2} \, \delta_{ij} \left( v_\pm^i \pm v_2^{ik} \p_k  \right)v_\pm^j + \mathfrak{U} \, .
\end{align}
\par 

Finally, the divergent Lagrangian \eqref{eq:divLag} generates a non-trivial integral condition of the form \cite{Kuipers:2023pzm}
\begin{equation}\label{eq:IntConst}
	\oint_\gamma\left(p_i^\pm v_\pm^i \pm \frac{1}{2} \, v_2^{ij} \p_j p_i^\pm \right) dt 
	=
	\pm \, \alpha \, \pi \, \ri \, k_i^i \, ,
\end{equation}
where $k\in\mathbb{Z}^{n\times n}$ is a matrix of winding numbers.
\par 

If we combine the Hamilton-Jacobi equations \eqref{eq:HamJac}, we find that the velocity field is a solution of the PDE
\begin{align}\label{stochHamJac}
	&\left[
	m \, \delta_{ij} \Big(
	\p_t 
	+ v_\pm^k \p_k 
	\pm \frac{\alpha}{2m} \delta^{kl} \p_l \p_k
	\Big)
	- q \, F_{ij} 
	\right] v_\pm^j  \nonumber\\
	&= 
	\pm \frac{\alpha\,q}{2\, m} \, \delta^{jk} \p_k F_{ij}
	- q \, \p_t A_i 
	- \p_i \mathfrak{U}
\end{align}
with field strength $F_{ij}=\p_i A_j - \p_j A_i$. It can easily be verified that this reproduces the classical Hamilton-Jacobi equations in the deterministic limit $\alpha\rightarrow0$. 
\par 

Eq.~\eqref{stochHamJac} can be solved for the velocity field $v_\pm$, which must be subjected to the integral constraint \eqref{eq:IntConst}. Then, the solution can be plugged into the It\^o equation \eqref{eq:Ito2}. This yields stochastic processes $X_\pm$, where $X_+$ defines the future directed evolution and $X_-$ the past directed evolution.
\par 

Alternatively, one can use the Hamilton-Jacobi equations \eqref{eq:HamJac} to derive the Kolmogorov backward equations for the probability densities of the processes $X_\pm$. This can be done by rewriting the Hamilton-Jacobi equations in the form
\begin{align}\label{eq:prediffusion}
	- 2 \, m \, \p_t S^\pm
	&=
	\p_i S^\pm \, \p^i S^\pm
	\pm \alpha \, \p_i\p^i S^\pm
	- 2\, q \, A^i \, \p_i S^\pm
	\nonumber\\
	&\quad
	\mp \alpha \, q \, \p_i A^i
	+ q^2 \, A_i A^i
	+ 2 \, m \, \mathfrak{U}\, .
\end{align}
Then, a straightforward calculation shows that the real wave function $\Psi:\mathcal{T}\times\M\rightarrow\R$, defined by
\begin{equation}\label{eq:Wavefunction}
	\Psi_\pm(x,t) = \exp\left(\pm \frac{S^\pm(x,t)}{\alpha} \right),
\end{equation}
is subjected to the diffusion equation
\begin{align}\label{eq:Diffusion}
	&\mp \alpha \frac{\p}{\p t} \Psi_\pm
	=\\
	&\left[
	\frac{\alpha^2}{2 m} 
	\left(\frac{\p}{\p x^i} \mp \frac{q}{\alpha} A_i \right)
	\left(\frac{\p}{\p x_i} \mp \frac{q}{\alpha} A^i \right)
	+ \mathfrak{U}
	\right] \Psi_\pm \, . \nonumber
\end{align}
This can be interpreted as the Kolmogorov backward equation for the process $X_\pm$ with respect to the $L^1$-norm, such that
\begin{equation}
	\rho_\pm(x,t) = \frac{|\Psi_\pm(x,t)|}{\int_{\R^n}|\Psi_\pm(y,t)| \, d^ny}
\end{equation}
is the probability density associated to the probability measure $\mu_\pm=\mathbb{P}\circ X_\pm^{-1}$ on $(\R^n,\mathcal{B}(\R^n))$ with Borel sigma algebra $\mathcal{B}(\R^n)$.
In addition, the integral constraint \eqref{eq:IntConst} implies the equivalence relation
\begin{equation}
	\tilde{S}^\pm \sim S^\pm \quad {\rm if} \quad \tilde{S}^\pm = S^\pm + \alpha \, \pi \, \ri \, k_i^i \, .  
\end{equation}
Due to this equivalence relation, eqs.~\eqref{eq:prediffusion} and \eqref{eq:Diffusion} are truly equivalent. Indeed, for any solution $\Psi_\pm$ of the diffusion equation \eqref{eq:Diffusion}, we can  construct an equivalence class $[\Psi_\pm]=\{+\Psi_\pm,-\Psi_\pm\}$ that defines a unique velocity field 
\begin{equation}
	v_\pm^i = \frac{\delta^{ij}}{m} \, \Big(\pm \alpha \, \p_j \ln [\Psi_\pm] - q \, A_j\Big),
\end{equation}
thus a stochastic process that solves the It\^o equation \eqref{eq:Ito2}.

\section{Complex Diffusion Theories}\label{sec:ComplexDiff}
In the previous section, we have studied the Wiener process, which describes Brownian motion, subjected to a vector and scalar potential, and found that the Hamilton-Jacobi equations for a Brownian particle are equivalent to the heat equation. We would now like to extend this analysis to complex diffusion constants $\alpha\in\mathbb{C}$. The main issue that arises in such an extension is that the real process $M$ with structure relation 
\begin{equation}\label{eq:genstructrel}
	d[M^a,M^b] = A^{ab} \, dt
\end{equation}
exists, if and only if $A$ is positive semi-definite, while for quantum theories one requires
\begin{equation}
	m \, d[M^a,M^b] = \ri \, \delta^{ab} \, dt \, .
\end{equation}
A similar issue arises in the relativistic extension of the Wiener process, which requires
\begin{equation}
	m \, d[M^a,M^b] = \alpha \, \eta^{ab} \, d\tau \, ,
\end{equation}
where $\tau$ labels the proper time. Due to the Minkowski signature, this fails to be positive semi-definite for any $\alpha\in\mathbb{C}$, thus for both quantum mechanics and Brownian motion.
\par 

The solution to this conundrum turns out to be rather simple.\footnote{We stress that the complex process studied here is different from the processes that were previously studied in stochastic mechanics \cite{Fenyes,Nelson,Nelson:1967,Nelson:1966sp,Guerra:1981ie}, including the complex formulation due to Pavon \cite{Pavon:1995April,Pavon:2000}. The advantage of this reformulation is twofold: (i) the complex process unifies quantum mechanics ($\alpha=\ri$) and Brownian motion ($\alpha=1$) in a single framework; (ii) the complex process correctly reproduces all aspects of quantum mechanics, whereas previous formulations failed to recover the correct multi-time correlations \cite{Nelson:2011}.}
Indeed, if we complexify the noise process $M$, we can study the structure relations
\begin{align}
	m \, d[M^a,M^b]_t 
	&=
	\alpha \, \delta^{ab} \, dt
	\, , \nonumber \\
	m \, d[M^a,\overline{M}{}^b]_t
	&=
	(|\alpha| + \gamma) \, \delta^{ab} \, dt
	\, , \nonumber\\
	m \, d[\overline{M}{}^a,\overline{M}{}^b]_t
	&=
	\overline{\alpha} \, \delta^{ab} \, dt \,.
\end{align}
Its decomposition $M=M_x+\ri \, M_y$ has a quadratic covariation given by
\begin{align}\label{eq:StructRelCompIn}
	&\begin{pmatrix}
		d[M_x^a,M_x^b]_t & d[M_x^a,M_y^b]_t\\
		d[M_y^a,M_x^b]_t & d[M_y^a,M_y^b]_t
	\end{pmatrix}
	\\
	&=\frac{\delta^{ab}}{2 \, m} 
	\begin{pmatrix}
		|\alpha|(1+\cos \phi) + \gamma & |\alpha| \, \sin \phi\\
		|\alpha| \, \sin \phi & |\alpha|(1 - \cos \phi) + \gamma
	\end{pmatrix}
	\, dt \, \nonumber ,
\end{align}
such that the process exists, if and only if this matrix is positive semi-definite, thus for any $\alpha=|\alpha|\,e^{\ri \phi}\in\mathbb{C}$ and $\gamma\in[0,\infty)$. The price we pay for its existence is that $M$ is now a complex process with $2n$ degrees of freedom instead of $n$, but we can ameliorate this by setting $\gamma=0$, such that 
\begin{align}
	d[M_x^a,M_x^b] &= \frac{|\alpha| \, (1+\cos\phi)}{2\, m} \, \delta^{ab}\, dt \, , \nonumber\\
	d[M_x^a,M_y^b] &= \frac{|\alpha| \, \sin \phi}{2\, m} \, \delta^{ab}\, dt \, ,\nonumber\\
	d[M_y^a,M_y^b] &= \frac{|\alpha| \, (1-\cos\phi)}{2\, m} \, \delta^{ab}\, dt\, .
\end{align}
It immediately follows that $M$ describes a real Wiener process on the hyperplane $e^{\frac{\ri \phi}{2}}\times\R^n\subset\mathbb{C}^n$.
\par 

The decomposition \eqref{DoobMeyer2} of the stochastic process now changes to 
\begin{equation}\label{DoobMeyer3}
	X^i_t = C^i_{\pm,t} + e^i_a(X_t) \, {\rm Re}[M^a_t] \, ,
\end{equation}
where the configuration space $\M=\R^n$ remains real, but the frames $F^\mathbb{C}=\mathbb{C}^n$ are complexified. Moreover, the quadratic variation of the complex process $M$ is characterized by the structure relation
\begin{equation}\label{eq:QVarComp}
	d[M^a,M^b]_t = \frac{\alpha}{m} \, \delta^{ab} \, dt.
\end{equation}
\par

Since only the process $M_x$ enters directly into the decomposition \eqref{DoobMeyer3}, it is tempting to ignore the process $M_y$. However, in general, this cannot be done, as the process $M_y$ enters the decomposition \eqref{DoobMeyer3} indirectly, due to its non-vanishing correlations with $M_x$. Only in the Brownian limit, $\alpha\in\R$, the processes $M_x$ and $M_y$ are uncorrelated, but they are maximally correlated in the quantum limit $\alpha\in\ri \times \R$.
\par 

Due to the complexification of the frames, the velocity fields $(v_\pm,\pm v_2)(x,t)$ become complex valued $(w_\pm,\pm w_2)(x,t)$ with decomposition $w=v+\ri \,u$. Therefore, the It\^o equation \eqref{eq:Ito2} is now given by
\begin{equation}\label{eq:Ito3}
	\begin{cases}
		d_\pm X^i_t &= {\rm Re} \left[ w_\pm^i(X_t,t) \, dt + \delta^i_a \, d_\pm M^a_t \right]\\
		d[M^a,M^b]_t &= \frac{\alpha}{m} \, \delta^{ab} \, dt
	\end{cases}
\end{equation}
and the Lagrangian becomes
\begin{equation}
	L^\pm(x,w_\pm,w_2,t) = L_0^\pm(x,w_\pm,w_2,t) \pm L_\infty(x,w_\circ) 
\end{equation}
with $L_0^\pm$ given by eq.~\eqref{eq:FinLag} and $L_\infty^{\pm}$ by eq.~\eqref{eq:divLag}. 
\par 

By replacing $v\rightarrow w$ it is straightforward to generalize all the results from section \ref{sec:EQM} to the complex processes with $\alpha\in\mathbb{C}$. In particular, $w_\pm(X_t,t)$ is still a solution of the complex PDE \eqref{stochHamJac} subjected to the constraint \eqref{eq:IntConst}. Thus, for any solution $w_\pm$ one can construct a real stochastic process $X$ by solving the It\^o equation \eqref{eq:Ito3}.
\par 

Moreover, one can construct diffusion equations \eqref{eq:Diffusion} with $\alpha\in\mathbb{C}$, such that complex wave functions $\Psi:\mathcal{T}\times\R^n\rightarrow\mathbb{C}$ that solve these diffusion equations can be related bijectively to the velocity field by
\begin{equation}
	w_\pm^i = \frac{\delta^{ij}}{m} \, \Big(\pm \alpha \, \p_j \ln [\Psi_\pm] - q \, A_j\Big)
\end{equation}
with $[\Psi_\pm]=\{+\Psi_\pm,-\Psi_\pm\}$. Hence, there is a one-to-one correspondence between solutions of the diffusion equation \eqref{eq:Diffusion} with $\alpha\in\mathbb{C}$ and stochastic processes that solve the It\^o equation \eqref{eq:Ito3}.
\par 

This stochastic theory is defined for any $\alpha\in\mathbb{C}$, thus, in particular, for the special cases $\alpha=1$, which describes a generalized Brownian motion, and $\alpha=\ri$, which reproduces quantum mechanics. Moreover, the real theory of section \ref{sec:EQM} that describes a standard Brownian motion is recovered by setting $\alpha\in(0,\infty)$ and imposing an initial condition $u_\pm(x,0)=0$ or terminal condition $u_\pm(x,T)=0$, such that $u_\pm(x,t)=0 \; \forall \, t\in\mathcal{T}$. 
\par 

We conclude this section by pointing out that by setting $\gamma=0$ in eq.~\eqref{eq:StructRelCompIn}, we have only partially succeeded in eliminating new degrees of freedom, as our theory contains a real position $x\in\R^n$, but complex velocities $w_\pm\in\mathbb{C}^n$. Therefore, we must provide a physical interpretation of the new degrees of freedom that are contained in the velocity $u_\pm$. 
\par 

The advantage of setting $\gamma=0$ resides in the fact that the process $M$ is restricted to the hyperplane $e^{\frac{\ri \phi}{2}}\times\R^n\subset\mathbb{C}^n$, which implies the constraint
\begin{align}
	u_+ \cos \frac{\phi}{2} - v_+ \sin \frac{\phi}{2} 
	=
	u_- \cos \frac{\phi}{2}  - v_- \sin \frac{\phi}{2}  \, .
\end{align}
Hence, there are only three independent velocity fields, and for $\phi\in(-\pi,\pi)$, we can choose these fields to be $v_+$, $v_-$ and $u_\circ=\frac{1}{2}(u_+ + u_-)$. 
\par 

The interpretation of the fields $v_\pm$ follows immediately from the theory of Brownian motion, where $v_\pm$ describe the drift velocities of a Brownian bridge, i.e. a Brownian motion with fixed endpoints. More precisely, $v_+$ describes the right limit of this drift velocity and $v_-$ the left limit. The fact that these limits are not equal to each other simply reflects the non-differentiability of the Wiener process. The physical interpretation of this fact is that the particle is kicked by the background field at every instant of time $t\in\mathcal{T}$, such that its velocity discontinuously changes from $v_-$ to $v_+$.
\par 

Using this physical picture of Brownian motion, we can also provide an interpretation for the field $u_\circ$. Indeed, if we interpret the background field as the continuum limit of a fluid consisting of a large number of microscopic particles, we can associate the velocity field $u_\circ(x,t)$ with the flow velocity of this fluid.

\section{Relativistic extensions}\label{eq:RLT}
Using the complex noise from previous section, we can also extend our stochastic theory from the Euclidean space $\R^n$ to the Minkowski space $\R^{n,1}$. This can be realized by imposing the structure relation
\begin{equation}
	d[M^\alpha,M^\beta]_\lambda = \alpha \, \varepsilon \, \eta^{\alpha\beta} \, d\lambda
\end{equation}  
with affine parameter\footnote{In non-relativistic theories the evolution parameter is the time $t$, while in relativistic theories the time $t$ is promoted to a coordinate $x^0=c\,t$. Instead, the evolution of relativistic dynamics is defined with respect to an arbitrary affine parameter $\lambda$. If $m>0$, this parameter reduces to the proper time $\tau$ of the particle after gauge fixing $\varepsilon$.} $\lambda$ and an auxiliary variable $\varepsilon$. This auxiliary variable must be gauge fixed in the equations of motion with the usual condition
\begin{itemize}
	\item $\varepsilon=m^{-1}$, if $m^2>0$, which fixes the affine parameter $\lambda=\tau$ to be the proper time;
	\item $\varepsilon=1$, if $m^2=0$, which normalizes the momentum such that $p_\mu=\eta_{\mu\nu}w^\nu + q A_\mu$ and fixes the dimension of the affine parameter to be $[\lambda]=T/M$;
	\item $\varepsilon=|m|^{-1}$, if $m^2<0$, which fixes the affine parameter $\lambda=s$ to be the proper length.
\end{itemize}
\par

In addition, in order to make the theory relativistic, we must impose a causality condition. In a classical theory, causality is defined by the statement that along any path $X$, the velocity field $v(X_\lambda)$ is everywhere time-like or null, such that 
\begin{equation}\label{eq:ClassCaus}
	\eta_{\mu\nu} v^\mu v^\nu(X_\lambda) \leq 0 \qquad \forall\, \lambda\in\mathcal{T} \, .
\end{equation}
In practice, this condition is imposed by fixing the rest mass $m\geq0$, such that causality is ensured by the energy-momentum relation
\begin{equation}\label{eq:EMClass}
	\eta_{\mu\nu} v^\mu v^\nu(X_\lambda) =  - \varepsilon^2 \, m^2 \quad \forall \,\lambda\in\mathcal{T} \, .
\end{equation}
\par 

In the stochastic theory, one can also fix the rest mass $m\geq0$, but the causality condition \eqref{eq:ClassCaus} is not well-defined, as this expression would require a distance measure for the individual sample paths. Such a norm on the path space is not available, but we can use the norm on the Lebesgue space $L_\mathcal{T}^2(\Omega)$ of all complex stochastic processes $Z:\mathcal{T} \times \Omega \rightarrow \mathbb{C}^{n+1}$. This norm is given by
\begin{align}\label{eq:L2Norm}
	||Z||^2 &= \E\left[\int_\mathcal{T} \eta_{\mu\nu} \, Z^\mu_\lambda \overline{Z}{}^\nu_\lambda \, d\lambda\right] 
	\\
	&= \int_\Omega \left[\int_\mathcal{T} \eta_{\mu\nu} \, Z^\mu_\lambda(\omega) \, \overline{Z}{}^\nu_\lambda(\omega) \, d\lambda\right] d\mathbb{P}(\omega)\, ,\nonumber
\end{align}
where $Z=X+\ri \, Y$ and $Y$ is an auxiliary process that solves the It\^o equation
\begin{equation}
	\begin{cases}
		d_\pm Y^\mu_\lambda &= {\rm Im} \left[ w_\pm^\mu(X_\lambda) \, d\lambda + e^\mu_\alpha(X_\lambda)\, d_\pm M^\alpha_\lambda \right]\\
		d[M^\alpha,M^\beta]_\lambda &= \alpha \, \varepsilon \, \eta^{\alpha\beta} \, d\lambda \, .
	\end{cases}
\end{equation}
\par 

The expectation value, which is an essential part of this norm, is inherited by the definitions of causality and by the energy-momentum relation, such that the classical conditions \eqref{eq:ClassCaus} and \eqref{eq:EMClass} are replaced by the stochastic conditions 
\begin{alignat}{2}
	\E\left[ \eta_{\mu\nu} w_\circ^\mu w_\circ^\nu(X_\lambda)  \right] &\leq 0 \quad &&\forall \lambda\in\mathcal{T} \\
	\E\left[ \eta_{\mu\nu} w_\circ^\mu w_\circ^\nu(X_\lambda)  \right] &=  - \varepsilon^2 \, m^2 \quad &&\forall \lambda\in\mathcal{T}  \label{eq:EMStoch}
\end{alignat}
with $w_\circ = \frac{1}{2}(w_+ + w_-)$ \cite{Kuipers:2023pzm}. 
\par 

The fact that the energy-momentum relation \eqref{eq:EMStoch} is imposed using an expectation value implies that for every $\omega\in\Omega$ the sample path $X(\omega):\mathcal{T}\rightarrow\M$ may violate causality, as long as the probabilistic average over all paths is causal. One can estimate \cite{Kuipers:2023pzm} that for any sample path such violations are likely to occur on scales smaller than the Compton wavelength, while such violations are exponentially suppressed on larger scales, i.e. for proper time intervals
\begin{equation}
	\Delta \tau \gtrsim  \frac{n \, (1+\cos\phi)}{2 \, m}\,.
\end{equation}
\par 

The sample paths of the stochastic process can be interpreted as the paths that one sums over in a path integral approach. Hence, the stochastic theory shows that these paths are continuous but neither differentiable nor causal. The non-differentiability of these path provides a physical interpretation of phase space non-commutativity \cite{Feynman:1948ur}. 
Similarly, the acausality of these paths provides an interpretation of the presence of negatively normed eigenstates in the spectrum of relativistic quantum theories. This stochastic interpretation is different from the usual canonical interpretation, where the causality violations are associated to the presence of anti-particles in a field theoretic reformulation of the theory. 
\par 

We emphasize that the two pictures are not in contradiction with each other, but should be regarded as complementary. The complex stochastic theory can be reformulated on a Fock space as is done in second quantization, cf. e.g. Ref.~\cite{Biane:2010}. In such a reformulation, one can adopt the usual interpretation, in which the different excitations of the background field are associated to particle number.
However, the existence of a stochastic process associated to the Klein-Gordon equation disproves the widespread belief, cf. e.g. Ref.~\cite{Peskin:1995ev}, that there does not exist a relativistic quantum theory associated to a single particle.\footnote{This is not the first work to point this out, as relativistic quantum theories for a single particle have been developed earlier in the literature, cf. e.g. Ref.~\cite{Reisenberger:2001pk}.}

\section{Extensions to manifolds}
The stochastic theory can also be generalized to the context of pseudo-Riemannian geometry. For more detail we refer to Ref.~\cite{Kuipers:2023pzm}. Here, we will only discuss the main difficulty that is encountered in this generalization: the stochastic It\^o Lagrangian \eqref{eq:FinLag} is not covariant, as the vector representation $v_\pm^\mu$ does not transform contravariantly under coordinate transformations.
\par 

This issue can be resolved \cite{Schwartz,Meyer,Emery,Huang:2022} by interpreting the vector representation $v_\pm^\mu$ as the first order part of a second order vector 
\begin{equation}
	v_\pm = v_\pm^\mu \p_\mu \pm \frac{1}{2} v_2^{\nu\rho} \p_{\nu\rho}\, ,
\end{equation}
where $\{\p_{\mu},\p_{\nu\rho}\}$ is the canonical basis of a second order tangent space. Using the affine connection, one can then construct covariant representations
\begin{align}
	\hat{v}_\pm^\mu &= v_\pm^{\mu} \pm \frac{1}{2} \, \Gamma^\mu_{\nu \rho } v_2^{\nu\rho} \nonumber\\
	\hat{v}_2^{\mu\nu} &= v_2^{\mu\nu} 
\end{align}
and
\begin{align}
	\hat{\p}_\mu &= \p_\mu \nonumber\\
	\hat{\p}_{\mu\nu} &= \p_{\mu\nu} - \Gamma^\rho_{\mu\nu} \p_\rho \, ,
\end{align}
such that the vector representation $(\hat{v}_\pm^{\mu},\pm \hat{v}_2^{\nu\rho})$ transforms contravariantly and the basis $(\hat{\p}_\mu,\pm \hat{\p}_{\nu\rho})$ covariantly. 
\par 

The generalization of the stochastic theory to pseudo-Riemannian geometry can now be achieved by consistently replacing the non-covariant vector representations $v_\pm^\mu$ by covariant representations $\hat{v}_\pm^\mu$, and by adding the Pauli-DeWitt term \cite{DeWitt:1957,Pauli:1973,Nelson,Kuipers:2023pzm} 
\begin{equation}
	\frac{1}{12\, \varepsilon} \mathcal{R}_{\mu\nu\rho\sigma} v_2^{\mu\rho} v_2^{\nu\sigma}
\end{equation}
to the Lagrangian.

\section{General stochastic theory}
We summarize the equations of motion of a stochastic particle on a pseudo-Riemannian manifold subjected to various potentials. We refer to Ref.~\cite{Kuipers:2023pzm} for a detailed derivation.

\subsection{Non-Relativistic case}
We consider a particle with mass $m>0$ on a $n$-dimensional Riemannian manifold $\M$. We assume that the particle is charged under a scalar potential $\mathfrak{U}$ and vector potential $A$ with charge $q\in\R$, and that the particle is subjected to a stochastic noise with diffusion constant $\alpha\in\mathbb{C}$.
\par 

The trajectory of this particle is given by a sample path of a stochastic process $X$ that solves the stochastic differential equation
\begin{equation}\label{eq:ItoNRLT}
	\begin{cases}
		d_\pm X^i_t &= {\rm Re} \left[ w_\pm^i(X_t,t) \, dt + e^i_a(X_t)\, d_\pm M^a_t \right] \\
		d[M^a,M^b]_t &= \frac{\alpha}{m} \, \delta^{ab} \, dt \, ,
	\end{cases}
\end{equation}
where the polyad $e^i_a$ is defined by the relation $g^{ij}=e^i_a e^j_b \delta^{ab}$. Moreover, the velocity field ${w_\pm^i=\hat{w}^i \mp \frac{\alpha}{2m}g^{jk}\Gamma^i_{jk}}$ is a solution of the equation
\begin{align}
	&\left[
	m \, g_{ij} \Big(
	\p_t 
	+ \hat{w}_\pm^k \nabla_k \Big)
	\pm \frac{\alpha}{2} \Big( g_{ij} \Box - \mathcal{R}_{ij}\Big)
	- q \, F_{ij} 
	\right] \hat{w}_\pm^j  \nonumber\\
	&=
	\frac{\alpha^2}{12\, m} \nabla_i \mathcal{R} 
	\pm \frac{\alpha\,q}{2\, m} \, \nabla^j F_{ij}
	- q \, \p_t A_i 
	- \nabla_i \mathfrak{U} \, ,
\end{align}
where $F_{ij}=\nabla_i A_j - \nabla_j A_i$ is the field strength and $\mathcal{R}_{ij}$ is the Ricci tensor. This solution may be multi-valued, as it is subjected to the non-trivial integral constraint 
\begin{equation}
	\oint_\gamma\left[ \Big(\hat{w}_\pm^i \pm \frac{\alpha}{2m} \nabla^i\Big) \Big(m \hat{w}_{\pm,i} + q A_i \Big) \right] dt
	=
	\alpha \, \pi \, \ri \, k_i^i \, .
\end{equation}
Furthermore, one can construct a wave function as in eq.~\eqref{eq:Wavefunction} that solves the diffusion equation
\begin{align}\label{eq:DiffusionNRLT}
	&\mp \alpha \frac{\p}{\p t} \Psi_\pm
	=\\
	&\left[
	\frac{\alpha^2}{2 m} 
	\left(\nabla_i \mp \frac{q}{\alpha} A_i \right)
	\left(\nabla^i \mp \frac{q}{\alpha} A^i \right)
	- \frac{\alpha^2}{6} \, \mathcal{R}
	+ \mathfrak{U}
	\right] \Psi_\pm \, .\nonumber
\end{align}
This may be interpreted as the Kolmogorov backward equation for the process $X_\pm$ with respect to the $L^2$-norm, such that
\begin{equation}
	\rho_\pm(x,t) = \frac{|\Psi_\pm(x,t)|^2}{\int_\M \sqrt{|g|} \, |\Psi_\pm(y,t)|^2 \, d^ny}
\end{equation}
is the probability density associated to the probability measure $\mu_\pm=\mathbb{P}\circ X_\pm^{-1}$ on $(\M,\mathcal{B}(\M))$.
Finally, for any solution of this diffusion equation, one can construct a stochastic process described by the It\^o equation \eqref{eq:ItoNRLT} with drift velocity
\begin{equation}
	\hat{w}_\pm^i = \frac{g^{ij}}{m} \, \Big(\pm \alpha \, \nabla_j \ln [\Psi_\pm] - q \, A_j\Big) \, .
\end{equation}

\subsection{Relativistic case}
We consider a particle with mass $m\geq0$ on a $(n+1)$-dimensional Lorentzian manifold $\M$. We assume that the particle is charged under a vector potential $A$ with charge $q\in\R$, and that the particle is subjected to a stochastic noise with diffusion constant $\alpha\in\mathbb{C}$.
\par 

The trajectory of this particle is given by a sample path of a stochastic process $X$ that solves the stochastic differential equation
\begin{equation}\label{eq:ItoRLT}
	\begin{cases}
		d_\pm X^\mu_\lambda &= {\rm Re} \left[ w_\pm^\mu(X_\lambda) \, d\lambda + e^\mu_\alpha(X_\lambda)\, d_\pm M^\alpha_\lambda \right]\\
		d[M^\alpha,M^\beta]_\lambda &= \alpha \, \varepsilon \, \eta^{\alpha\beta} \, d\lambda \, .
	\end{cases}
\end{equation}
where the polyad $e^\mu_\alpha$ is defined by the relation ${g^{\mu\nu}=e^\mu_\alpha e^\nu_\beta \eta^{\alpha\beta}}$. Moreover, the velocity field ${w_\pm^\mu=\hat{w}^\mu \mp \frac{\alpha \varepsilon}{2} g^{\nu\rho}\Gamma^\mu_{\nu\rho}}$ is a solution of the equation
\begin{align}
	&\left[
	g_{\mu\nu} \hat{w}_\pm^\rho \nabla_\rho 
	\pm \frac{\alpha \, \varepsilon}{2} \Big( g_{\mu\nu} \Box - \mathcal{R}_{\mu\nu}\Big)
	- \varepsilon \, q \, F_{\mu\nu} 
	\right] \hat{w}_\pm^\nu  \nonumber\\
	&=
	\frac{\alpha^2 \varepsilon^2}{12} \nabla_\mu \mathcal{R} 
	\pm \frac{\alpha\, \varepsilon^2\, q}{2} \, \nabla^\nu F_{\mu\nu}\, ,
\end{align}
and is subjected to the integral constraint 
\begin{equation}
	\oint_\gamma\left[ \Big(\hat{w}_\pm^\mu \pm \frac{\alpha}{2m} \nabla^\mu\Big) \Big(m \hat{w}_{\pm,\mu} + q A_\mu \Big) \right] dt
	=
	\alpha \, \pi \, \ri \, k_\mu^\mu \, .
\end{equation}
Furthermore, the wave function can be defined as in eq.~\eqref{eq:Wavefunction} and is given by\footnote{$\Psi$ is the solution of a complex diffusion equation on the space $\M\times\mathcal{T}$, where the dynamics is measured with respect to the affine parameter $\lambda$. Due to the reparameterization invariance of the relativistic theory, this diffusion equation can be solved by separation of variables, yielding eq.~\eqref{eq:PsiRLT}, where $\Phi$ satisfies the Klein-Gordon equation \eqref{eq:DiffusionRLT}.}
\begin{equation}\label{eq:PsiRLT}
	\Psi_\pm(x,\varepsilon,\lambda) = \Phi_\pm(x) \, \exp\left(\pm \frac{\varepsilon \, m^2}{2 \, \alpha} \, \lambda\right),
\end{equation}
where $\Phi_\pm$ is a solution of the wave equation
\begin{align}\label{eq:DiffusionRLT}
	\left[ 
	\Big(\nabla_\mu \mp \frac{q}{\alpha} A_\mu \Big)
	\Big(\nabla^\mu \mp \frac{q}{\alpha} A^\mu \Big)
	- \frac{\mathcal{R}}{6} 
	+ \frac{m^2}{\alpha^2}
	\right] \Phi_\pm = 0 \, .
\end{align}
This may be interpreted as the Kolmogorov backward equation for the process $X_\pm$ with respect to the $L^2$-norm, such that
\begin{equation}
	\rho_\pm(x,\lambda) = \frac{|\Psi_\pm(x,\lambda)|^2}{\int_\M \sqrt{|g|} \, |\Psi_\pm(y,\lambda)|^2 \, d^{n+1}y}
\end{equation}
is the probability density associated to the probability measure $\mu_\pm=\mathbb{P}\circ X_\pm^{-1}$ on $(\M,\mathcal{B}(\M))$.
Finally, for any solution of this diffusion equation, one can construct a stochastic process described by the It\^o equation \eqref{eq:ItoRLT} with drift velocity
\begin{equation}
	\hat{w}_\pm^\mu = \varepsilon \, g^{\mu\nu} \, \Big(\pm \alpha \, \nabla_\nu \ln [\Psi_\pm] - q \, A_\nu\Big) \, .
\end{equation}

\section{Conclusion}
We have established an equivalence between solutions of the complex diffusion equation \eqref{eq:DiffEq} and stochastic processes that are subjected to a stochastic variational principle and a structure relation as in eq.~\eqref{eq:QVarComp}. For $\alpha\in \R$, this equivalence is well-known, as it is established by the Feynman-Kac theorem \cite{FKac}. 
The novelty of our result is the construction of a complex noise, which, in combination with the stochastic variational principle, allows to extend this equivalence to Lorentzian quantum theories. 
\par

This idea is very close in spirit to the Wick rotation that is often employed in the study of quantum theories. In this Euclidean approach, one maps a (relativistic) quantum theory onto a Euclidean theory. Then, by the Feynman-Kac theorem, this Euclidean theory can be studied using the standard Wiener process. Here, we have taken the opposite approach: we have extended the equivalence that is established by the Feynman-Kac theorem by allowing for complex degrees of freedom, such that the Wiener process itself can be Wick rotated. Due to this extension, the original (relativistic) quantum theory can be studied without performing a Wick rotation on this theory.
\par 

In addition, the results show that both quantum mechanics and Brownian motion of the single spinless particle can be regarded as special cases within a large class of complex diffusion theories. This fact has important consequences for the foundations of quantum mechanics, as it imposes an interpretation of quantum mechanics that is similar to that of Brownian motion. 
\par 

Due to the complexification of the theory, we find that both the particle itself and the particles in the background field are subjected to a Wiener process. The major difference between a Brownian theory and a quantum theory is that a Brownian particle is not correlated with the background, while a quantum particle is maximally correlated with the background. This maximal correlation ensures that quantum theories are unitary and sources the wavelike behavior of quantum mechanics, as opposed to the dissipative behavior of Brownian motion.

\section*{Acknowledgements}
This research was carried out in the frame of Programme STAR Plus, financially supported  by UniNA and Compagnia di San Paolo. 

\section*{Competing Interests}
The authors have no competing interests to declare that are relevant to the content of this article.

\section*{Data Availability}
Data sharing not applicable to this article as no datasets were generated or analysed during the current study.

\end{multicols}
\end{document}